# Optimal covering of rectangular grid graphs with tours of constrained length

**Sergey Bereg** ✉
Department of Computer Science, University of Texas at Dallas, USA

**Jesús Capitán** ✉
Multi-robot & Control Systems group, University of Seville, Spain

**José Miguel Díaz Bañez** ✉
Department of Applied Mathematics, University of Seville, Spain

**José Manuel Higes López** ✉
Department of Applied Mathematics, University of Seville, Spain

**Miguel Angel Pérez Cutiño**[a] ✉
Department of Applied Mathematics, University of Seville, Spain
Virtualmechanics S.L, Seville, Spain

**Vanesa Sánchez Canales** ✉
Department of Applied Mathematics, University of Seville, Spain

[a] Corresponding author

**Inmaculada Ventura** ✉
Department of Applied Mathematics, University of Seville, Spain

―― **Abstract** ――――――――――――――――――――――――――――――――

Given a rectangular grid graph with a special vertex at a corner called base station, we study the problem of covering the vertices of the entire graph with tours that start and end at the base station and whose lengths do not exceed a given threshold, while minimizing a quality measure. We consider two objective functions: minimizing the number of tours and minimizing the sum of their lengths. We present an algorithm that computes the optimal solution for both objectives in linear time with respect to the grid size.

**2012 ACM Subject Classification** Mathematics of computing → Combinatorial optimization; Theory of computation → Packing and covering problems

**Funding** This work has been partially supported by grants PID2020-114154RB-I00, TED2021-129182B-I00 and DIN2020-011317 funded by MCIN/AEI/10.13039/501100011033 and the European Union NextGenerationEU/PRTR.

## 1 Introduction

Let $G$ be a rectangular grid graph, that is, a grid graph defined by an $a \times b$ rectangle such that all its interior faces are unit squares (i.e., the graph contains no holes). The vertex at the bottom left corner of $G$ is denoted as the base station $\mathcal{B}$. A tour is a walk starting and ending at $\mathcal{B}$ whose length, measured with the Manhattan metric, is bounded by $L > 0$. Our target problems are to compute a set of tours covering all vertices of $G$ so that: **(1)** the number of tours is minimized, or **(2)** the total length (the sum of the lengths of all tours) is minimized. We denote these problems as the **Min-Tours** and **Min-Length Problem**, respectively, and they are of practical interest for drone-based applications where the battery of the robot limits its flying period. Our case study is inspired by solar power plants inspection, as their topologies suggest to consider discretizations to grid points, and they are typically inspected by drones that move according to the Manhattan distance; see Figure 1.

Our problems are related with the family of Vehicle Routing Problems (VRP) [22] that fall within the framework of the TSP-type problems. The Traveling Salesman Problem (TSP)

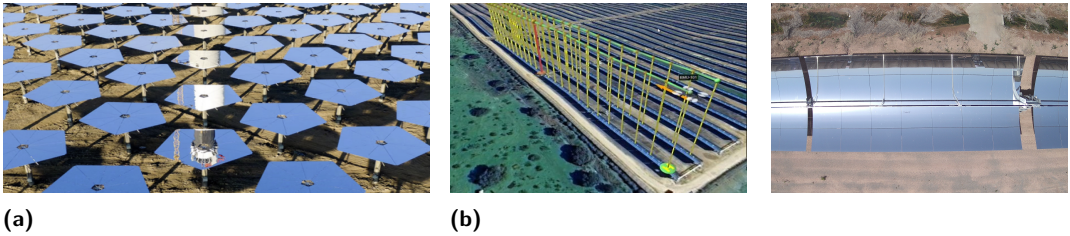

**(a)** **(b)**

■ **Figure 1** Example of solar power plants. (a) Stellio Heliostat delivers a solution to deploy high-efficiency solar power generation with grid-connected facilities. (b) Drone inspection route based on discrete points in Parabolic Trough plants (left), where the objective is to capture one image per solar collector element (right).

on solid grid graphs, that is, without holes, is one of the long-standing open problems in the prominent list *The Open Problems Project*[1] (TOPP). As a consequence, finding a set of length-constrained tours of minimal total length covering a graph remains open in general solid grids. This paper addresses the question of whether that problem is polynomially solvable in rectangular grid graphs, for the case in which all tours starts at a corner. To the best of our knowledge, this question has not been formally addressed in the existing literature.

A useful property is that minimizing the total length of the set of covering tours is equivalent to minimizing the number of repetitions on the vertices of $G$; see Section 2. Therefore, when solving the Min-Length Problem, we focus our attention on minimizing the total number of repeats. Although several strategies can be designed to obtain a covering of $G$, proving optimality is not straightforward even when the problems are defined over rectangular grid graphs. For instance, in Figure 2a, a natural greedy algorithm, which generates a set of tours with each tour having the maximum allowable length, fails to achieve an optimal solution for the Min-Length Problem, illustrated in Figure 2b.

An interesting result of our research is the connection between the Min-Tours and Min-Length Problems; see the second and third contributions listed at the end of this section. This connection is achieved by analyzing the following auxiliary problem, named **Range Level Problem**, which holds independent theoretical interest.

**Range Level Problem (RLP):** *Define the level $i \in \mathbb{N}$ of $G$ as the set of points at Manhattan distance $i$ from $\mathcal{B}$. Determine the maximum level $i$ such that for every set of tours $T$ covering $G$ with minimum number of repeats, each tour $t \in T$ reaches $i$.*

Let $n$ be the total number of vertices of $G$, and $k_{\min}$ the minimum number of tours required to cover the grid $G$. Our main contributions can be summarized as:

1. $k_{\min}$ can be found in $O(1)$ time.

2. The minimum number of repeats required to cover $G$ is at least $2k_{\min}^2 - 2k_{\min} - 1$, and at most $2k_{\min}^2 - 2k_{\min} + 1$.

3. In $O(n)$ time we can obtain a set of tours $T$ covering $G$ such that: (1) $|T| = k_{\min}$, and (2) the number of repetitions of $T$ in $G$ is minimum.

---

[1] http://cs.smith.edu/~orourke/TOPP/

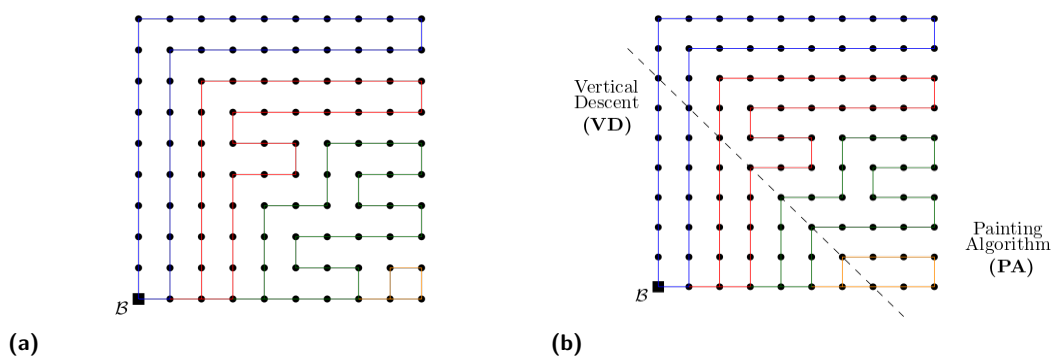

**Figure 2** A rectangular grid of $10 \times 10$. Each color defines a different tour with length at most 36. The connection to each tour with $\mathcal{B}$ is omitted for visualization purposes. (a) Example of a greedy strategy with total length 128. (b) Covering obtained from our algorithms with a total length of 124. Notice that the minimum number of tours, $k_{\min}$, required to cover the grid follows $k_{\min} = 4 > \lceil \frac{100}{36} \rceil$.

## 1.1 Related work

The most notable combinatorial optimization problems have been studied on grid graphs. The Hamiltonian Cycle Problem (HCP) has been proved NP-hard on general grid graphs [11], but it can be solved in polynomial time for solid grid graphs and quad-quad graphs [23]. Other authors have considered special topological structures such as grid graphs whose shapes are similar to alphabet characters. The so-called L-alphabet, C-alphabet, and O-alphabet grid graphs were considered as special cases of grid graphs that have Hamiltonian cycles [20]. Linear-time algorithms for finding Hamiltonian paths and cycles in a rectangular grid graph with one or two rectangular holes have been given [12, 14]. A similar computational time is obtained for truncated rectangular grid graphs [13]. For other classes of grid graphs, e.g., triangular and hexagonal grids, it is known that the HCP is NP-complete in the general case but decidable in polynomial time for solid triangular grids [2]. Indeed, the complexity for solid hexagonal grids is still an open problem. In the case of the TSP, the problem remains open for solid grid graphs, and some approximation algorithms are known on general grid graphs [4, 17]. Moreover, Arkin et al. [3] gave a $\frac{6}{5}$-approximation algorithm for the TSP problem in grid graphs without holes and a 1.325-approximation algorithm for grids containing holes. Grids with forbidden neighborhoods have also been considered [9]. Last, finding the shortest tour is polynomially solvable when restricted to thin grid graphs, i.e., grid graphs that do not contain an induced $2 \times 2$ square [1].

For rectangular grid graphs, different covering problems have been considered. Motivated by hypergraphs visualization, van Goethem et al. [24] studied the problem of covering a rectangular grid with colored connected polygons. More general problems on partitioning a set of points via non-crossing structures have been considered later [7]. Another related problem is the Shortest $k$-disjoint Paths, that looks for pairwise vertex-disjoint paths between $k$ specified pairs of vertices (terminals) within the graph so that the total length is minimized [16]. It has been proved that, for the case of a grid, the problem admits a fixed parameter tractable algorithm parameterized by the number of terminals [16]. It is well known that the $k$-disjoint Paths Problem is NP-complete in the case of infinite rectangular grids [15].

There is a vast body of literature related to finding a set of tours covering a general graph. The Multiple Traveling Salesman Problem (MTSP) is a generalization of the TSP that better reflects the needs in real applications. Exact formulations and heuristic approaches are presented in [5], and a solution with two salesmen in rectangular grids is obtained in [10]. For a comprehensive survey on this problem, see [6]. MTSP is a relaxation of the Capacitated MTSP (CMTSP) [19], where each vehicle or salesman has a limited fuel or energy capacity. Another related and celebrated problem is the Chinese Postman Problem (CPP). The multiple-vehicle extension is the $k$-Chinese Postman Problem ($k$-CPP): compute a set of $k$ tours so that (1) each route begins and ends at a common node, (2) each edge in $G$ must be traversed by exactly one tour, and (3) the total distance traveled is minimized. These problems are NP-hard in general [21], but with suitable restrictions on the graph, polynomial-time algorithms have been developed. For instance, if the graph is undirected, both the CPP and $k$-CPP are polynomial-time solvable [8, 18].

Our problems share similarities with the previously discussed ones but differ in key aspects. Unlike the CMTSP, we allow repetitions both in the vertices and edges of the grid; unlike the $k$-CPP, our problem does not require complete edge coverage, and the value of $k$ is not part of the input. The primary objective of the present work is to analyze the complexity of our target problems, that is, the Min-Tours and the Min-Length Problem, which are defined in the context of rectangular grid graphs.

## 1.2 Organization

The remainder of the paper is structured as follows. Our target problems are formally defined in Section 2; while in Section 3 the main algorithms used in our solution are described. Sections 4, 5, and 6 present the solutions to the Min-Tours, the Range Level and the Min-Length problems, respectively. Finally, we conclude in Section 7 with some open questions.

## 2 Problems Definition

Given a rectangular grid graph $G$, a special vertex or base station $\mathcal{B}$ located in the lower left corner of $G$, and an integer $L > 0$, we aim to compute a set of tours that cover the entire grid $G$. We consider that the length of each tour is constrained by $L$, and they must start from and return to $\mathcal{B}$. Since the length of any tour in $G$ is even, we assume that $L$ is even. Without loss of generality, we assume that the number of columns $n_c$ is less than or equal to the number of rows $n_r$ in $G$. Finally, for any graph $H$, we define $V(H)$ as the set of vertices of $H$.

▶ **Definition 1.** *Let $\ell(p)$ denote the length of the walk $p$. For a given $L > 0$, we say that a walk $p$ is a tour if it is a cycle (not necessarily simple) containing $\mathcal{B}$ with $\ell(p) \leq L$. Consequently, we define the length of a set of tours $T$ by $\ell(T) = \sum_{t \in T} \ell(t)$.*

Our objective is to solve two independent minimization problems:

**Min-Tours Problem (MTP).** Given a rectangular grid graph $G$ with a base station $\mathcal{B}$ in its bottom left corner, and an integer $L > 0$, find a set of tours $T$ covering $G$ with minimum number of tours $|T|$. If such $T$ exists, we denote $k_{\min} = |T|$.

**Min-Length Problem.** Given a rectangular grid graph $G$ with a base station $\mathcal{B}$ in its bottom left corner, and an integer $L > 0$, find a set of tours $T$ covering $G$ such that $\ell(T)$ is minimum.

The following lemma proves that in a rectangular grid graph, minimizing the total length of the tours is equivalent to minimize the total number of repetitions of the tours, that is, the total number of times each vertex is traversed by all tours except for its first traversal. Note that each vertex is traversed when it has an incoming and an outgoing edge in the tour; hence for any set of $k$ tours the total number of repeats at $\mathcal{B}$ is always $k-1$.

▶ **Lemma 2.** *For a given set of tours $T$ covering $G$, minimizing the total length of $T$ is equivalent to minimizing the total number of repeats.*

**Proof.** Consider $T$ as a set of tours covering $G$ with total length $\ell(T)$ and with $r$ repetitions. Clearly, $|V(G)| = \ell(T) - r$ and the lemma follows. ◀

Consequently, we focus on solving the following problem:

**Min-Repetitions Problem (MRP).** Given an integer $L > 0$ and a rectangular grid graph $G$ with a base station $\mathcal{B}$ in its bottom left corner, let $r(v,T)$ denote the number of repetitions of a set of tours $T$ in vertex $v$, i.e., the total number of times that vertex $v \in V(G)$ is traversed by all tours in $T$ minus 1. The problem is to find a set of tours $T^*$ covering $G$ minimizing $\sum_{v \in G} r(v, T^*)$.

When proposing a solution for MTP and MRP, the following definition and problem arise in a natural way.

▶ **Definition 3.** *The level $i$ of a rectangular grid graph $G$ is the set of vertices in $G$ that are at the same Manhattan distance $i$ from the base station $\mathcal{B} \in G$. Note that at level $i$, there are $i+1$ points if $i \leq n_c - 1$.*

**Range Level Problem (RLP):** Given a rectangular grid graph $G$, the base station $\mathcal{B}$ and $L > 0$, determine the maximum level $i$ such that for every set of tours $T$ covering $G$ with minimum number of repeats, each tour $t \in T$ reaches $i$.

As stated in Section 1, the Range Level Problem will be used as an auxiliary problem to show the connection between MTP and MRP.

## 3 The Algorithms

In this section, we provide formal definitions for the proposed algorithms. First, we introduce some additional notation.

▶ **Definition 4.** *Let $A_i$ denote the grid graph obtained by removing from $G$ all vertices located below level $i$. Here, level $i$ of $G$ is referred to as the baseline of $A_i$. Vertices at the baseline of $A_i$ are denoted from left to right by $t_1, t_2, \ldots, t_{i+1}$.*

Our strategy divides the grid into two parts: $A_{2r-1}$ and $G \setminus A_{2r-1}$ for every $r \in \{1, \ldots, \lceil \frac{n_c}{2} \rceil\}$. We then focus on computing an optimal set of $r$ pairwise disjoint walks covering $A_{2r-1}$, where each walk in the set starts from and returns to the baseline of $A_{2r-1}$. In order to define some constraints for these walks, we introduce the following definitions. First, we introduce a family of staircase grid graphs with 0, 1, or 2 steps of height 2. The elements of this family are important substructures used at each step of our algorithm.

▶ **Definition 5.** *Let $\mathcal{S}$ be the family of grid graphs represented in Figure 3, composed by the following grid graphs:*
- $\mathcal{R}(a,b)$: *a rectangular grid with $a$ columns and $b$ rows (see Figure 3a).*
- $\mathcal{S}_1(a,b,c)$: *a grid graph obtained from $\mathcal{R}(a,b)$, $b \geq 2$, by removing its subgraph $\mathcal{R}(a-c,2)$ from the upper left corner, for any $c < a$ (see Figure 3b).*

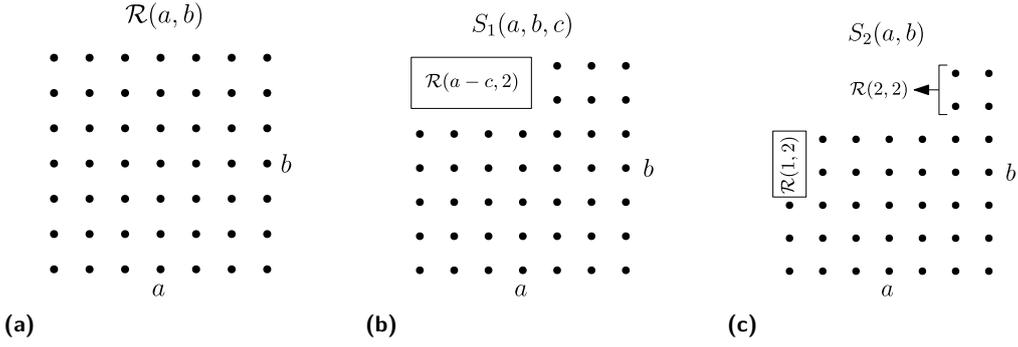

**Figure 3** General grids contained in the family $\mathcal{S}$, according to Definition 5.

- $S_2(a, b)$: *a grid graph obtained from $\mathcal{R}(a, b-2)$, $a, b \geq 3$, by removing its subgraph $\mathcal{R}(1, 2)$ from the upper left corner and adding a graph $\mathcal{R}(2, 2)$ right aligned above its right top corner (see Figure 3c).*

When making a reference to a particular subfamily of the grids contained in $\mathcal{S}$, we use the expression $\mathcal{R}(\cdot)$, $S_1(\cdot)$, or $S_2(\cdot)$ for convenience. Similarly, the expression $S_1(\cdot, \cdot, 1)$ will denote any grid in the subfamily of $S_1(\cdot)$ with $c = 1$.

▶ **Definition 6.** *For any grid graph $H$, let $z$ be the number of vertices in the bottom row of $H$. When $z \geq 2$ we define a new grid graph, $AddRow(H)$, as the grid graph obtained by adding one row with $z$ points at the bottom of $H$, followed by the removal of the left-most point of this newly added row.*

▶ **Definition 7.** *Given $U \in \mathcal{S}$ with $a > 1$ columns, and an integer $l$ being at least the perimeter of $U$, let $A = AddRow(U)$ and $p$ be a walk over $A$ such that $\ell(p) \leq l$. We define $A(p)$ as the portion of $A$ that is covered by $p$, and $R$ as its complementary, i.e., $R = A \setminus A(p)$. In addition, we say that $p$ is $\mathcal{S}$-maximal if all of the following conditions hold:*
1. $R \in \mathcal{S}$.
2. *Assuming the points in a row of $A$ are ordered from left to right, for any row there exists a point $r$ such that all points $r'$ of such row with $r' \leq r$ belong to $A(p)$, and all the remaining points in such row, if any, are in $R$.*
3. *Assuming the points in a column of $A$ are ordered from top to bottom, for any column there exists a point $c$ such that all points $c'$ of such column with $c' \leq c$ belong to $A(p)$, and all the remaining points in such column, if any, are in $R$.*
4. *Either $\ell(p) = l$ or $R \in \{\emptyset, \mathcal{R}(a-2, 1), S_1(a-2, 3, 2)\}$.*

To compute $r$ pairwise disjoint walks covering $A_{2r-1}$ that start and end on its baseline, we perform an iterative procedure. In every step of the algorithm, a non-covered portion $U \in \mathcal{S}$ of $A_{2r-1}$ is selected and partially or totally covered using $\mathcal{S}$-maximal walks. When $A_{2r-1}$ is fully covered, each walk is transformed into a tour starting and ending at $\mathcal{B}$ by vertically descending from the baseline of $A_{2r-1}$ to the bottom row of $G$, and then using a horizontal line to connect with $\mathcal{B}$, if needed. We denote this last process as **Vertical Descent (VD)**. This combined strategy covers completely the grid; see Figure 2b. If the lengths of the walks defined in $A_{2r-1}$ are properly upper bounded by $L - 2(2r-1)$, then the resulting tours meet the length constraint. Notice that if we get $r$ pairwise disjoint walks covering $A_{2r-1}$, then all repetitions of the **VD** process appear in the bottom line of the grid.

■ **Algorithm 3.1 $\mathcal{U}$-Covering**

**Require:** $A = AddRow(U)$ with $a > 1$ columns for some $U \in \mathcal{S} \setminus \mathcal{S}_1(\cdot, \cdot, 1)$, and $l \in \mathbb{N}$ being at least the perimeter of $A$ minus 2. Let $t_1$ be the bottom left vertex of $U$, $t_2$ the bottom left vertex of $A$, and $t_a$ the top right vertex of $A$. Assume the rows of $A$ are numbered from the bottom starting in 1.

**Ensure:** A walk $p$ in $A$ such that $p$ is $\mathcal{S}$-maximal.

**1.** Starting from $t_1$, connect it with $t_a$ using the left and top border of $A$; let $p_1$ be the resulting walk. Then go down one unit from $t_a$, and connect this vertex with $t_2$ using the top and left border of $A \setminus A(p_1)$; let $p_2$ be the resulting walk. Finally, consider $p = p_1 \bigcup p_2$. If $\ell(p) = l$, then stop.

**2.** Let $j$ be the upper row of $A$ with a set of $n$ points not covered by $p$. If $\ell(p) < l$ and $j \geq 2$, extend $p$ in row $j$ and $j-1$ from left to right using $\min(n, \frac{l-\ell(p)}{2})$ points of each row.

**3.** If $\ell(p) = l$, and only one point in row $j$ is left uncovered, then remove from $p$ the last covered point in row $j$ and in row $j-1$, then cover two additional points of the next two rows, $j-2$ and $j-3$, if possible.

**4.** If $A$ is fully covered, or the uncovered portion of $A$ is its bottom row except for $t_2$, or $\ell(p) = l$, or $\ell(p) = l-2$ and $j \leq 3$, then stop.

**5.** Go to step 2.

■ **Algorithm 3.2 One Tour $\mathcal{U}$-Covering** ($OT\mathcal{U}$-Covering)

**Require:** $A = AddRow(U)$ for some $U \in \mathcal{S} \setminus \mathcal{S}_1(\cdot, \cdot, 1)$, and $l \in \mathbb{N}$ being at least the area of $A$.

**Ensure:** A full covering of $A$.

**1.** Execute Step 1 of $\mathcal{U}$-Covering; see Algorithm 3.1.

**2.** Execute Step 2 of $\mathcal{U}$-Covering; see Algorithm 3.1.

**3.** Check if $A$ is fully covered, or if the uncovered portion of $A$ is a rectangle with one row. If the second condition applies, then finish the covering of $A$ using a comb; refer to Figure 4 for a detailed description. If any of the two conditions apply, then stop.

**4.** Go to step 2.

## 3.1 The Painting Algorithm

In this section, we describe an optimal method for covering $A_{2r-1}$ with $r$ walks, which we call **Painting Algorithm (PA)**. At each step, **PA** divides the uncovered portion of the grid into two zones: one to be covered and one to be left untouched. Interestingly, the portion to be covered is always a grid $A = AddRow(U)$ for some $U \in \mathcal{S}$. Therefore, we design two auxiliary procedures to be used by **PA**. The first one aims to connect the bottom left vertex of $U$, denoted as $t_1$, with the bottom left vertex of $A$, denoted as $t_2$, using an $\mathcal{S}$-maximal walk; see Figures 5b-d. We term this procedure as $\mathcal{U}$-**Covering**; see Algorithm 3.1. When $A$ can be covered with one tour of length at most $l$, we add some modifications to the $\mathcal{U}$-**Covering** strategy. We refer to this procedure as **One Tour $\mathcal{U}$-Covering** ($OT\mathcal{U}$-Covering) which is outlined in Algorithm 3.2.

We are now ready to describe **PA**. The algorithm divides $A_{2r-1}$ into two portions, as shown in Figure 5(b): 1) a zone $AddRow(U_1)$ containing $t_1$, $t_2$ and their corresponding rows in $A_{2r-1}$; 2) a trapezoid $\delta_1 = A_{2r-1} \setminus AddRow(U_1)$. At step $i > 1$ of the algorithm, two points ($t_{2i-1}$ and $t_{2i}$) in the baseline of $A_{2r-1}$ will be connected with an $\mathcal{S}$-maximal walk.

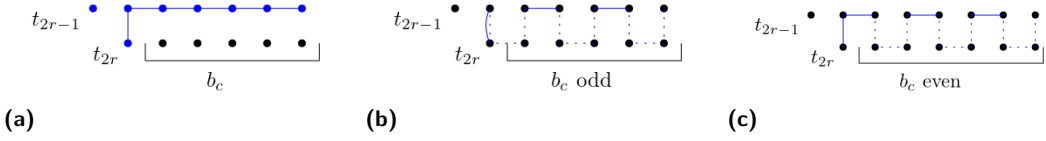

**(a)** **(b)** **(c)**

**Figure 4** Final step in OT$\mathcal{U}$-Covering, when the grid graph has an odd number of rows. (a) State of the covering (indicated with the blue color) after applying $\mathcal{U}$-Covering. (b) Modification of the walk created with $\mathcal{U}$-Covering when the number of columns is odd. Solid lines are used for edges in the original walk, while dashed lines are new edges. (c) The case of an even number of columns.

**Algorithm 3.3 Painting Algorithm (PA)**

---

**Require:** $G = \mathcal{R}(a,b)$, an integer $1 \leq r \leq \lceil \frac{a}{2} \rceil$, and $l \in \mathbb{N}$.
**Ensure:** $A_{2r-1}$ is fully covered with $r$ walks (if possible) and minimum number of repeats.
    $P \leftarrow \emptyset$
    $C \leftarrow$ all vertices of $A_{2r-1}$ above the row at which $t_1$ is located.
    **for** $i \in [1, \ldots, r-1]$ **do**
        $U_i \leftarrow C \bigcup$ row of $t_{2i-1}$
        $U' \leftarrow AddRow(U_i)$
        $p_i \leftarrow$ The resulting walk when applying $\mathcal{U}$-Covering over $U'$ with length at most $l$
        $C \leftarrow U_i \setminus p_i$
        $P = P \bigcup \{p_i\}$
    **end for**
    $U_r \leftarrow C \bigcup$ row of $t_{2r-1}$
    $U' \leftarrow AddRow(U_r)$
    $p_r \leftarrow$ The resulting walk when applying $OT\mathcal{U}$-Covering over $U'$ with length at most $l$
    **return** $P \bigcup \{p_r\}$

---

Moreover, $U_i$ will be the union of the uncovered points of $U_{i-1}$ and the points in the row that contains $t_{2i-1}$; and $\delta_i$ will be the trapezoid $A_{2r-1} \setminus \bigcup_{1 \leq j \leq i} AddRow(U_j)$; see Figure 5 for a complete example. The procedure is outlined in Algorithm 3.3.

**PA** will be used to prove the following theorems, which allow us to solve MRP and MTP. The details of the proofs are provided in the next section.

▶ **Theorem 8.** *Let $P$ be a set of $r$ walks, each of length at most $l$, that start and end on the baseline of $A_{2r-1}$ and collectively cover $A_{2r-1}$. Then, there exists a set $P'$ of pairwise disjoint walks, each of length at most $l$, starting and ending on the baseline of $A_{2r-1}$ such that:*
    *(i) $P'$ covers $A_{2r-1}$,*
    *(ii) at most one walk in $P'$ contains a single repetition, and*
    *(iii) the ending and starting points of every walk are different.*

▶ **Theorem 9.** *PA is optimal for covering $A_{2r-1}$ with $r$ walks and minimum number of repeats. Moreover, it runs in $O(|V(A_{2r-1})|)$ time.*

## 3.2 Proving Theorems 8 and 9

We state some technical results before proving Theorems 8 and 9.

▶ **Lemma 10.** *$\mathcal{U}$-Covering produces an $\mathcal{S}$-maximal walk $p$ with $O(\ell(p))$ operations.*

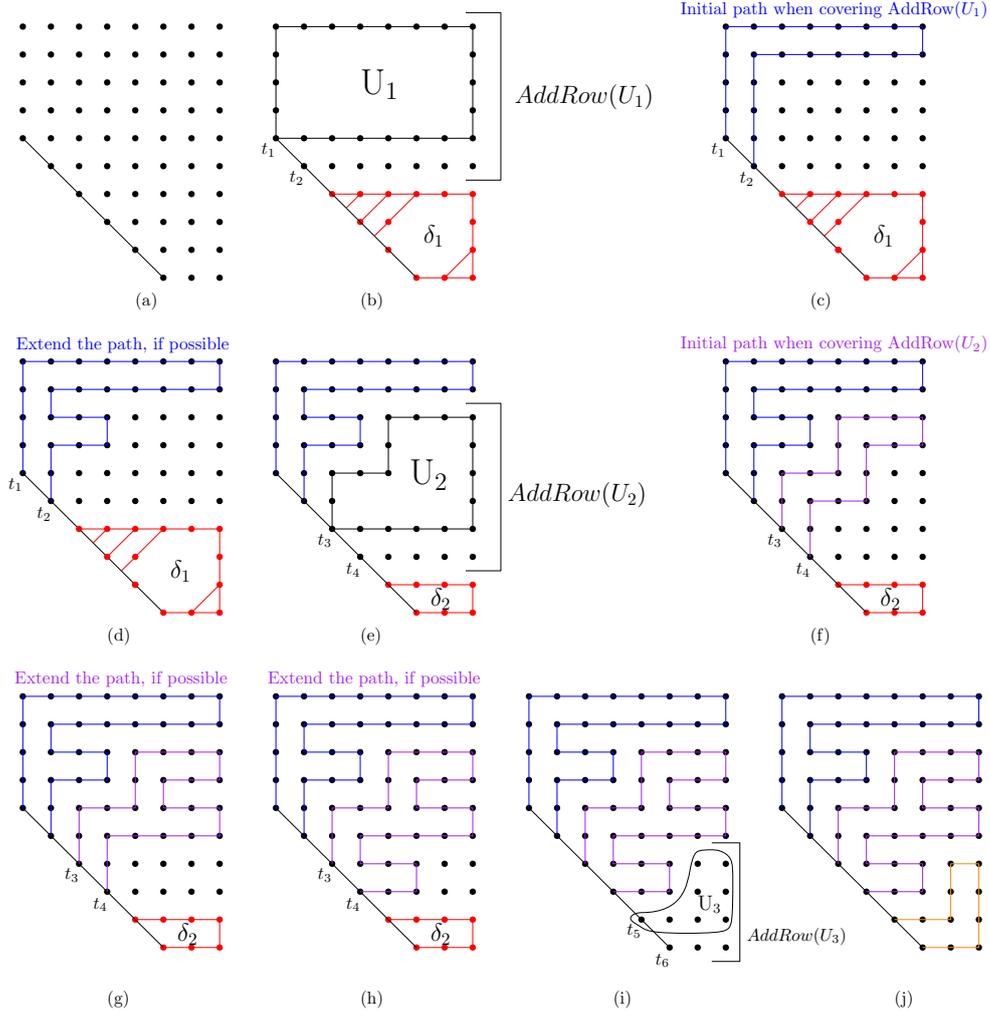

**Figure 5** Example of Painting Algorithm for covering $A_{2r-1}$ with $l = 26$. (a) $A_5$ with 10 rows and 8 columns; (b) initial zone division in the algorithm; (c)-(d) creating the first walk iteratively; (e) first walk (in blue) and second zone division; (f)-(g)-(h) creating the second path iteratively; (i) first two walks (in blue and purple) and third zone division; (j) final covering.

**Proof.** We use the notation of Algorithm 3.1 and Definition 7. Clearly, step 1 of $\mathcal{U}$-Covering produces a walk $p$ satisfying conditions 1, 2 and 3 from Definition 7. Then, if $\ell(p) = l$ the algorithm ends and $p$ is $\mathcal{S}$-maximal. Otherwise, the upper row $j$ of $A = AddRow(U)$ with some points not covered by $p$ is selected. Because $R = A \setminus A(p)$ belongs to $\mathcal{S}$, $p$ can be extended in $A$ using points in row $j$ and $j-1$. In other words, $p$ can be extended in the two upper rows of $R$, which always contains the same number of points. This ensures that the uncovered portion of $A$ belongs to $\mathcal{S}$ even after increasing $p$. The process is repeated iteratively within the algorithm until some condition at step 4 is satisfied. Notice that conditions at step 4 of $\mathcal{U}$-Covering are equivalent to the fourth statement in Definition 7; hence $p$ is $\mathcal{S}$-maximal. Moreover, the total number of operations is $O(\ell(p))$, as the decision of adding an uncovered point of $A$ to $p$ in every step of the algorithm can be solved in constant time. ◀

▶ **Lemma 11.** *Let $p$ be the walk resulting from applying $\mathcal{U}$-Covering over $A = AddRow(U)$*

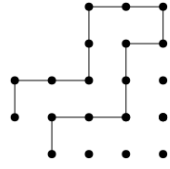

**Figure 6** Example of the walk produced at step 1 of $\mathcal{U}$-Covering. The walk could leave one point uncovered in two rows when $l+2$ is exactly the perimeter of the grid.

*for some $U \in \mathcal{S}$ that is a valid input of the algorithm. Then the number of rows fully covered by $p$ is always even.*

**Proof.** Since $U \in \mathcal{S}$, step 1 of the algorithm ensures that the two upper rows of $A$ are always fully covered. In any other step where $p$ is modified, the number of covered rows is increased in pairs (for instance, row $j$ and row $j-1$ in step 2). Then the Lemma holds. ◀

▶ **Lemma 12.** *Assume $A = AddRow(U)$ can be covered with one walk $w$ of length at most $l \in \mathbb{N}$, for some $U \in \mathcal{S} \setminus R(1, \cdot) \setminus S_1(\cdot, \cdot, 1)$, where $w$ starts at the bottom left vertex of $U$ and ends at the bottom left vertex of $A$. Then, $OT\mathcal{U}$-Covering finds a walk $p$ of length at most $l$ that covers $A$ with minimum number of repeats.*

**Proof.** Let $a, b$ be the number of columns and rows of $A$, respectively. As $U \in \mathcal{S} \setminus R(1, \cdot) \setminus S_1(\cdot, \cdot, 1)$, then no repeats are introduced in the first step of the algorithm. Two stop conditions are considered during the execution of $OT\mathcal{U}$-Covering: either $A$ is fully covered using $\mathcal{U}$-Covering, or the walks defined in Figure 4 are combined with the result of $\mathcal{U}$-Covering. In the first scenario, $b$ is even and $p$ makes no repeats; then $OT\mathcal{U}$-Covering is optimal. In the second case, $b$ is odd. If $a$ is even, then $p$ makes no repeats (see Figure 4c) and $OT\mathcal{U}$-Covering is optimal. Finally, when $a$ and $b$ are odd $OT\mathcal{U}$-Covering produces exactly one repetition. We can demonstrate this is optimal. Assume $A$ can be covered without repeats. Let $b_1$ be $b$ if $U$ is a rectangle, $b-2$ if $U \in S_1(\cdot)$, or $b-4$ if $U \in S_2(\cdot)$, and consider $R = \mathcal{R}(a, b_1)$. Clearly, if $A$ can be covered with a Hamiltonian path connecting the bottom left vertex of $U$ with the bottom left vertex of $A$, then $R$ can be covered with a Hamiltonian cycle; hence, $a \times b_1$ is even. However, $b_1 \equiv b \mod 2$, hence $a \times b_1$ is odd, a contradiction. Therefore, at least one repeat is required to cover $A$. ◀

▶ **Lemma 13.** *Let $P$ be any set of $r$ walks covering $A_{2r-1}$, with each walk starting and ending on its baseline. If $|V(G)|$ is odd, then there is at least one repetition of $P$ in $A_{2r-1}$.*

**Proof.** We will prove that if the number of repetitions of $P$ in $A_{2r-1}$ is 0 then, $|V(G)|$ is even. Assume the points on the baseline are numbered in increasing order from left to right. Since walks in $P$ start and end on the baseline of $A_{2r-1}$, they are disjoint if and only if the following condition is met: for any walk starting in point $i$ and ending in point $j > i$ of the baseline it follows that any walk starting at $i < z < j$ ends at $z < w < j$. Thus, $P$ can be easily transformed into disjoint cycles in $G$, which can be joined to create a Hamiltonian cycle. Therefore, $|V(G)|$ is even and the lemma holds. ◀

**Proof of Theorem 8.** By construction, **PA** produces disjoint walks satisfying conditions (i) and (iii) of Theorem 8. The second condition of Theorem 8 requires at most one repetition for the union of the walks.

At step $i$ of **PA**, let $p_i$ be the walk produced by the $\mathcal{U}$-Covering algorithm in its step 1, and $P_i$ be the resulting walk from the algorithm[2]. At step 1 of **PA**, $U_1$ is a rectangle of width 2, 4 or bigger. Therefore, $p_1$ never leaves one point in two consecutive rows uncovered. Interestingly, this is not the case for $p_j$ when $1 < j < r$, see for instance Figure 6. However, it is easy to see that $\ell(p_j) \geq \ell(p_{j+1})+4$; therefore, if $p_j$ leaves one point in two consecutive rows uncovered, they will be covered by $P_j$. Notice that $p_j$ will never leave one point uncovered in more than 2 rows unless $U_j$ is a rectangle of width 3, which is not possible for $j < r$. Moreover, at least two points per row will be left uncovered by $P_j$. Thus, we can assume that after applying $\mathcal{U}$-Covering in the first $r-1$ steps of **PA**, no repeats are produced.

We now consider the step $r$ of **PA**. Let $A' = AddRow(U_r)$ be the uncovered portion of $A_{2r-1}$ at step $r$, and let $a, b$ be the number of columns and rows of $A'$, respectively. If $r = 1$ or during the execution of the first $r-1$ steps **PA** produces walks of maximum length, then $l$ is enough to cover $|V(A')|$ points. Because of the restrictions of $\mathcal{U}$-Covering, and because $A'$ has 2, 4 or more columns, $U_r \notin R(1, \cdot) \cup S_1(\cdot, \cdot, 1)$; hence, according to Lemma 12 $A'$ can be covered optimally using $OT\mathcal{U}$-Covering. In the other case, $r \geq 2$ (implying $n_c \geq 4$) and some walks in **PA** are not of maximum length. Then, by the stop conditions of $\mathcal{U}$-Covering, $A'$ has at most 5 rows; in such case $U_r = S_1(a, 4, 2)$. The size of $A'$ is then exactly $n_c - (2r-1) + 2(n_c - 2r + 2) + 4 = 3n_c - 6r + 9$. Since $l$ is at least $2n_r + 2n_c - 4r + 2$ and $n_c \leq n_r$, then $l + 1 - |V(A')| \geq n_c + 2r - 7 > 0$. Therefore, $l$ is enough to cover $|V(A')|$ points and, as in the previous case, $A'$ can be optimally covered with $OT\mathcal{U}$-Covering. ◀

**Proof of Theorem 9.** From the proof of Theorem 8 and Lemma 13, it follows that **PA** yields an optimal set for covering $A_{2r-1}$ with $r$ walks and minimum number of repeats. In terms of complexity, **PA** applies $\mathcal{U}$-Covering $r-1$ times and $OT\mathcal{U}$-Covering one time. By Lemma 10, the total number of operations performed by **PA** is $O(\sum_{i=1}^{r} l_i)$, where $l_i$ is the length of the walk created with $\mathcal{U}$-Covering at step $i < r$ and, with the $OT\mathcal{U}$-Covering at step $i = r$. Clearly, $\sum_{i=1}^{r} l_i = O(|V(A_{2r-1})|)$ and the result holds. ◀

## 4 Minimum number of tours

The Painting Algorithm (**PA**) presented in Section 3.1 will be important to solve MTP. First, consider the following algorithm to obtain the value of $k_{\min}$:

▪ **Algorithm 4.1** *Kmin-Finder*
___
**Require:** $G = \mathcal{R}(n_c, n_r)$ and $L \in \mathbb{N}$, with $L \geq 2(n_c + n_r - 2)$.
**Ensure:** The value of $k_{\min}$ tours.
  1. If $L > n_c \cdot n_r$, or $L = n_c \cdot n_r$ and $|V(G)|$ is even, return 1.
  2. From $i = 2$ to $\lceil \frac{n_c}{2} \rceil - 1$, apply **PA** to $A_{2i-1}$ with length $l = L - 2(2i - 1)$.
  3. Stop at the first level $2j - 1$ such that $A_{2j-1}$ can be covered using **PA** with $j > 1$ tours, and $A_{2j-3}$ cannot be covered using **PA** with $j - 1$ tours. Return j.
  4. If none of the above holds, return $\lceil \frac{n_c}{2} \rceil$.
___

▶ **Lemma 14.** *Kmin-Finder finds the value of $k_{min}$.*

**Proof.** If $k_{\min} = 1$, then ***Kmin-Finder*** finds its value at step 1. Otherwise, assume ***Kmin-Finder*** stops at level $2j - 1$ such that $A_{2j-3}$ cannot be covered using **PA** with $j - 1$ walks of

---
[2] In Figure 5, $p_1$ is illustrated by subfigure (c), and $P_1$ by subfigure (d); $p_2$ is illustrated by subfigure (f), and $P_2$ by subfigure (h); $p_3$ and $P_3$ are the same, and they are illustrated in subfigure (j).

maximum length $l = L - 2(2j - 3)$; see step 3. Then, according to Theorem 8, no set of $j - 1$ walks covers $A_{2j-3}$ with length upper bounded by $l$, hence no set of $j - 1$ tours of length at most $L$ covers $G$. As **Kmin-Finder** finds a set of $j$ walks covering $A_{2j-1}$, which can be transformed into $j$ valid tours covering $G$ using **VD**, then $k_{\min} = j$. Finally, if condition at step 3 never applies, then clearly $k_{\min} = \lceil \frac{n_c}{2} \rceil$. ◀

Due to Theorem 9, **Kmin-Finder** runs in $O(k_{\min}|V(G)|)$. However, this time complexity can be greatly improved. Notice that the value of $k_{\min}$ cannot be obtained by simply dividing $\frac{|V(G)|}{L}$; see a counterexample in Figure 2. However, since **PA** is optimal for covering $A_{2r-1}$, the following decision question is solvable in constant time: *can $A_{2r-1}$ be covered with $r$ walks of length at most $l$?* $A_{2r-1}$ can be covered with 0 repeats iff $|V(G)|$ is even; see Lemma 13 and Theorems 8 and 9. Then, for a given $l$ and $r$, if $k_{\min} > 1$, we can establish the following formula to obtain its value:

$$k_{\min} = \begin{cases} \min\{r : |V(A_{2r-1})| \leq r(L - 2(2r-1) + 1)\} & \text{if } |V(G)| \text{ is even,} \\ \min\{r : |V(A_{2r-1})| + 1 \leq r(L - 2(2r-1) + 1)\} & \text{if } |V(G)| \text{ is odd.} \end{cases} \quad (1)$$

Solving the above equation, we get the following theorem:

▶ **Theorem 15.** *The minimum number of tours of length at most $L$ required to cover a rectangular grid graph $G$ can be obtained in constant time, and is given by:*

$$k_{min} = \begin{cases} \left\lceil \frac{L+2-\sqrt{-8n_c n_r + L^2 + 4L + 4}}{4} \right\rceil, & |V(G)| \text{ is even} \\ \\ \left\lceil \frac{L+2-\sqrt{-8n_c n_r + L^2 + 4L - 4}}{4} \right\rceil, & |V(G)| \text{ is odd} \end{cases} \quad (2)$$

As $k_{\min}$ can be obtained in constant time, we can achieve a set of tours covering $G$ using **PA+VD** over $A_{2k_{\min}-1}$. We will refer to this set of tours using the notation $\mathbf{PA}_{k_{\min}}\mathbf{+VD}$. Then, we get:

▶ **Theorem 16.** $\mathbf{PA}_{k_{min}}\mathbf{+VD}$ *solves MTP in $O(|V(G)|)$.*

## 5     Range Level Problem

To solve MRP, we first establish key connections between MTP and MRP. These connections arise from the analysis of RLP. In this section, we present a tight lower bound for RLP, and begin by stating several important results.

▶ **Lemma 17.** *Let $T$ be a set with $k_{min}$ tours covering $G$, and let $T' \subset T$ be the set of tours in $T$ whose points are all located on or below level $2k_{min} - 3$. Then, $|T'| \leq 1$.*

**Proof.** Assume that $|T'| \geq 2$. The cardinality of $T \setminus T'$ is then at most $k_{\min} - 2$. Consider the tour $t''$ that goes from the base point $\mathcal{B}$ to one extreme of level $2k_{\min} - 3$ and covers the level with an stair walk (see Figure 7). Notice that if level $2k_{\min} - 3$ does not exceed the main diagonal then $l(t'') = 4(2k_{\min} - 3) \leq L$. Now, the set of tours $T'' = T \setminus T' \cup \{t''\}$ covers $A_{2k_{\min}-3}$. Moreover, if we consider just the portions of these tours that are contained within $A_{2k_{\min}-3}$, they form at most $k_{\min} - 1$ walks that cover $A_{2k_{\min}-3}$ and begin and end on its baseline. Therefore, the inequality in Equation 1 holds for $k_{\min} - 1$, a contradiction. ◀

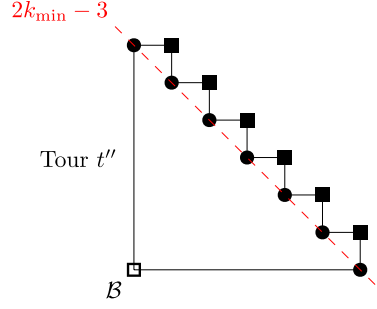

**Figure 7** t" can be added to $T \setminus T'$ to get a covering of $A_{2k_{\min}-3}$ with at most $k_{\min} - 1$ random walks that satisfy the conditions of Theorem 8. Level $2k_{\min} - 3$ is indicated with a dashed red line.

▶ **Lemma 18.** *Let $T$ be a set with $k_{min}$ tours covering $G$, and let $T' \subset T$ be the set of tours in $T$ whose points are all located on or below level $2k_{min} - 3$. If $|T'| = 1$, then $r(v, T) = 0$ for every $v \in V(G)$ above level $2k_{min} - 3$, i.e., there are no repetitions above level $2k_{min} - 3$. Moreover, for any $t \in T \setminus T'$, the number of points covered by $t$ above level $2k_{min} - 3$ is at least 8.*

**Proof.** Let $t'$ be the tour in $T'$ and $d$ the number of points covered by $t'$ on the baseline of $A_{2k_{\min}-3}$, such that no other tour in $T$ covers these points. Notice that there are $k_{\min} - 1$ tours with at least one point above level $2k_{\min} - 3$, and $2k_{\min} - 2 - d$ available points on the baseline of $A_{2k_{\min}-3}$. Therefore, the number of repeats on the baseline of $A_{2k_{\min}-3}$ is at least $2(k_{\min} - 1) - (2k_{\min} - 2 - d) = d$. Then, the maximum number of points that $T$ can cover in $A_{2k_{\min}-3}$ is $(k_{\min} - 1)(L - 2(2k_{\min} - 3) + 1) - d - r$, where $r$ is the number of repeats of $T$ at the vertices of $A_{2k_{\min}-3}$ that are not on the baseline. Clearly, the following inequality holds:

$$(k_{\min} - 1)(L - 2(2k_{\min} - 3) + 1) - d - r \geq |V(A_{2k_{\min}-3})| - d$$

Notice that if $|V(G)|$ is even or $r > 0$, then Equation 1 is satisfied for $k_{\min} - 1$, a contradiction. Therefore, $r = 0$.

Now assume that there exists a tour $t \in T \setminus T'$ whose points above level $2k_{\min} - 3$ are at most 7. Then, we can separate this tour from the other $k_{\min} - 2$ and write the previous inequality as $(k_{\min} - 2)(L - 2(2k_{\min} - 3) + 1) + 7 \geq |V(A_{2k_{\min}-3})|$, which implies that Equation 1 is satisfied for $k_{\min} - 1$, a contradiction. Thus, the lemma holds. ◀

▶ **Lemma 19.** *Let $P$ be a set with $k_{min} - 1$ walks defined over $A_{2k_{min}-3}$ such that each walk $p \in P$ starts and ends on the baseline of $A_{2k_{min}-3}$, and satisfies $3 \leq \ell(p) \leq L - 2(2k_{min} - 3)$. If $P$ makes no repetitions above the baseline of $A_{2k_{min}-3}$, then there exists a set $P'$ of $k_{min} - 1$ walks such that every point on the baseline of $A_{2k_{min}-3}$ is the starting or ending point of some walk $p' \in P'$, and each $p'$ is exactly $p$ for some $p \in P$, excluding their starting and ending points.*

**Proof.** Let $S(P, p)$ be the vertex set containing the starting and ending points of a walk $p \in P$, and $S(P) = \bigcup_{p \in P} S(P, p)$. In addition, let $F(P)$ be the vertices on the baseline of $A_{2k_{\min}-3}$ that are not a starting or ending point of any walk in $P$. We say a vertex $s \in S(P)$ is a swap point if $S(P, p) = \{s\}$ for some $p \in P$, or $s \in S(P, p_1)$ and $s \in S(P, p_2)$ for some $\{p_1, p_2\} \subseteq P$ with $p_1 \neq p_2$. Let $d = 2k_{\min} - 2 - |S(P)|$ be the number of swap points in $S(P)$. Clearly, $d$ matches the number of elements in $F(P)$, because $F(P) + S(P) = 2k_{\min} - 2$. Considering the premises of this lemma, the scenarios depicted in Figure 8 are not possible. Therefore, it is easy to see that to the left of a swap point only vertical edges are used as

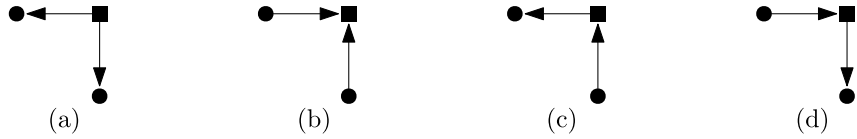

**Figure 8** Possible scenarios for vertices (represented as solid squares) adjacent to vertices on the baseline of $A_{2k_{\min}-3}$ (represented as solid circles) that are either a starting or an ending point of some walk covering $A_{2k_{\min}-3}$. The arrows indicate the direction of the walk. (a) and (b) are not possible if no repetitions are allowed at the square vertex. This is also the case for (c) and (d), unless the walk passing through the square vertex has exactly length 2.

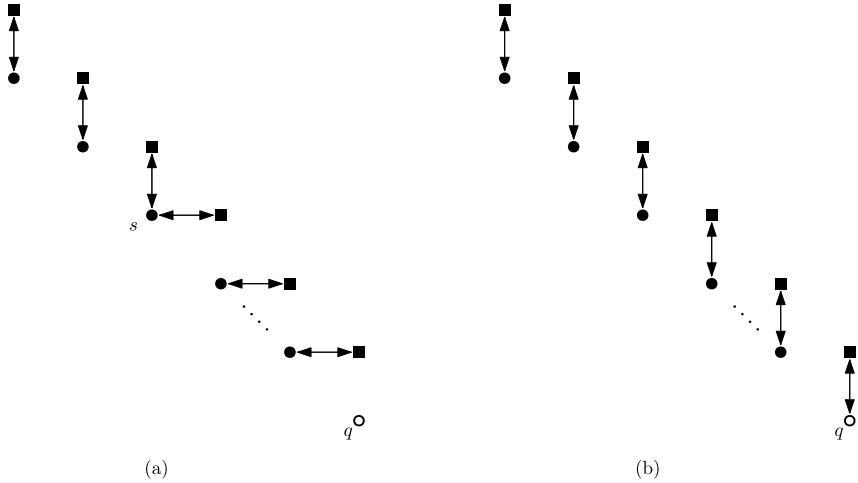

**Figure 9** The diagram shows how to move the starting and ending points of some walks covering $A_{2k_{\min}-3}$ to a target point $q$, according to the definitions of Lemma 19. (a) Initial configuration, (b) modified paths. The figures only include the final portion of the path indicated with arrows in two directions. Notice that to the left (top) of the swap point $s$ all arrows have to be vertical, while to the right (bottom) of $s$ all arrows have to be horizontal, if the scenarios depicted in Figure 8 are not possible.

starting or ending edges, until a point $q \in F(P)$ is reached. Equivalently, to the right of the swap points only horizontal edges are used. This indicates that between two consecutive swap points there is always a vertex $q \in F(P)$. Given $P$, we obtain a set of walks $P'$ by *moving* the swap point until a point $q \in F(P)$ is reached. The procedure for moving the swap point $s$ to the vertex $q$ is depicted in Figure 9. As $d = |F(P)|$, after a finite number of steps, all swap points in $S(P)$ can be moved to obtain the set of walks $P'$ where $S(P') = 2k_{\min} - 2$, finishing the proof. ◀

▶ **Lemma 20.** *Let $T$ be a set with $k_{min}$ tours covering $G$. Then, every $t \in T$ contains at least one point above level $2k_{min} - 3$.*

**Proof.** Let $T' \subset T$ be the tours in $T$ that contain all its points on or below level $2k_{\min} - 3$. By Lemma 17, $|T'| \leq 1$. Assume $|T'| = 1$. According to Lemma 18, $T$ makes no repeats above level $2k_{\min} - 3$. Consider $P$ as the set of walks resulting from removing the points below level $2k_{\min} - 3$ to the set of tours in $T \setminus T'$. Clearly $|P| = k_{\min} - 1$, $A_{2k_{\min}-3}$ is covered by $P$ except for some vertices on its baseline, and every $p \in P$ has length at least 8; see Lemma 18. Then by Lemma 19, there exists a set of $k_{\min} - 1$ walks of length at most $L - 2(2k_{\min} - 3)$ that fully covers $A_{2k_{\min}-3}$. Therefore, the inequality in Equation 1 is satisfied for $k_{\min} - 1$, a contradiction; hence $|T'| = 0$ and this lemma holds. ◀

Now, we introduce some important results that relate MRP with MTP. This relationship is key to derive our lower bound for the Range Level Problem.

▶ **Lemma 21.** *Let $T$ be a set of $k$ tours defined in $G$ such that $m$ tours in $T$ contain at least one point above some level $j \leq 2m - 1$. Then, the total number of repeats of $T$ in $G$ is at least $2mj + k - \frac{(j+1)(j+2)}{2}$.*

**Proof.** By the Pigeonhole Principle, we know that the total number of repeats of $m$ tours in any level $0 < i \leq 2m - 1$ is at least $2m - (i + 1)$. In addition, for $i = 0$ the number of repeats is exactly $k - 1$. Adding up to level $j$, we obtain the formula. ◀

▶ **Lemma 22.** *Let $T$ be the set of tours covering $G$ obtained with $\mathbf{PA}_{k_{\min}} + \mathbf{VD}$. Then, the total number of repetitions of $T$ in $G$ is at most $2k_{min}^2 - 2k_{min} + 1$.*

**Proof.** Notice that the number of repeats obtained with $\mathbf{VD}$ below level $2k_{\min} - 1$ is $2k_{\min}^2 - 2k_{\min}$. From Theorem 8, we get the (possible) additional repeat from $\mathbf{PA}_{k_{\min}}$. ◀

▶ **Theorem 23.** *For any set of tours covering $G$, the minimum number $r$ of repeats satisfies:*

$$2k_{min}^2 - 2k_{min} - 1 \leq r \leq 2k_{min}^2 - 2k_{min} + 1$$

**Proof.** The upper bound is obtained directly from Lemma 22. For the lower bound, notice that any set of tours will contain at least $k_{\min}$ tours such that each tour contains at least one point above level $2k_{\min} - 3$ by Lemma 20. Then, applying Lemma 21 for $m = k_{\min}$ and $j = 2k_{\min} - 3$, we get the formula. ◀

▶ **Theorem 24.** *Let $T$ be a set of tours covering $G$ with minimum number of repetitions; then $|T| = k_{min}$.*

**Proof.** (By contradiction) Clearly $|T| \geq k_{\min}$; assume $|T| > k_{\min}$. There exist then a set of tours $T' \subset T$ with $|T'| = k_{\min}$ such that each tour in $T'$ contains at least one point above level $2k_{\min} - 3$; see Lemma 20. Using Lemma 21 with $m = k_{\min}$ and $j = 2k_{\min} - 3$, the total number of repeats of $T'$ in $G$ is at least $2k_{\min}^2 - 2k_{\min} - 1$. Let $t \in T$ be a tour such that $t \notin T'$. As $T$ is an optimal solution of MRP and $|T| \geq 2$, $\ell(t) \geq 4$. Now, either there is a free point in level 1 of $G$ after performing all the tours in $T'$, or both points in level 1 are covered. In the first scenario, the minimum number of repetitions of $T'$ in $G$ is at least $2k_{\min}^2 - 2k_{\min}$, and the number of repeats of $t$ in $G$ is at least 2. On the second case, the number of repeats of $t$ in $G$ is at least 3; hence the total number of repetitions of $T' \cup t$ is at least $2k_{\min}^2 - 2k_{\min} + 2$. Then by Theorem 23, the total number of repetitions obtained with $T$ is not minimal. ◀

Using Theorem 24 and Lemma 20, the solution to the Range Level Problem can be readily obtained:

▶ **Theorem 25.** *The solution $z$ to the **RLP** problem satisfies that $2k_{min} - 2 \leq z$.*

## 6 Minimum total length

In this section, we start by analyzing the conditions for which $\mathbf{PA}_{k_{\min}}+\mathbf{VD}$ is optimal for solving MRP, that is, for minimizing the total number of repetitions. Then, we extend **PA** to consider two additional cases by modifying $\mathcal{OTU}$-Covering. With these tools, we demonstrate that **PA+VD** (including the modified versions of **PA**) finds an optimal solution of MRP for any $G$ and $L$. From this point forward, when referring to optimal solutions, we will consider a set of tours covering $G$ with minimum number of repetitions, unless otherwise specified. In what follows, it will be demonstrated that:

- If $k_{\min} < \lceil \frac{n_c}{2} \rceil$:
  - $\mathbf{PA}_{k_{\min}} + \mathbf{VD}$ is optimal when $|V(G)|$ is even, or when there exist an optimal solution $T$ to MRP such that every tour in $T$ traverses level $2k_{\min} - 1$.
  - If $\mathbf{PA}_{k_{\min}} + \mathbf{VD}$ is not optimal, then $|V(G)|$ is odd and exactly 1 tour does not reach level $2k_{\min} - 1$. We modify the final tour of $\mathbf{PA}_{k_{\min}}$ to achieve an optimal solution in this scenario.
- If $k_{\min} = \lceil \frac{n_c}{2} \rceil$:
  - If $n_c$ is even, then $\mathbf{PA}_{k_{\min}} + \mathbf{VD}$ is optimal.
  - If $n_c$ is odd, we modify the final tour of $\mathbf{PA}_{k_{\min}}$ to achieve an optimal solution in this scenario.

First, we state an important result.

▶ **Lemma 26.** *Let $d$ be the number of tours in an optimal solution $T$ to MRP that does not reach the level $2k_{min} - 1$. If $d > 0$, then $d = 1$.*

**Proof.** (By contradiction). Assume $d \geq 2$, and for simplicity, let $k = k_{\min}$. Then $A_{2k-1}$ is covered with $T$ using $k - 2$ tours, resulting in: $|V(A_{2k-1})| \leq (k-2)(L - 2(2k-1) + 1)$. Using $|V(A_{2k-1})| = |V(A_{2k-3})| - (2k-1) - (2k-2)$, and the previous inequality we get $|V(A_{2k-3})| \leq (k-1)(L - 2(2k-3) + 1) - L + 4k - 2$. We know that $k \leq \frac{\min(n_c, n_r)}{2}$ and $L \geq 4\min(n_c, n_r) - 4$; therefore $-L + 4k - 2 < -1$. Replacing on the previous formula we get:

$$|V(A_{2(k-1)-1})| + 1 \leq (k-1)(L - 2(2(k-1)-1) + 1).$$

Then by Equation 1 the value of $k_{\min}$ is $k - 1$, a contradiction. ◀

### 6.1 The case $k_{\min} < \lceil \frac{n_c}{2} \rceil$

In this Section we assume $k_{\min} < \lceil \frac{n_c}{2} \rceil$ unless otherwise specified. We analyze the optimallity conditions for $\mathbf{PA}_{k_{\min}} + \mathbf{VD}$ in this scenario.

▶ **Theorem 27.** *If there exists a set of tours $T$ covering $G$ with minimum number of repeats such that every tour in $T$ traverses level $2k_{min} - 1$, then $\mathbf{PA}_{k_{min}} + \mathbf{VD}$ produces an optimal solution for MRP.*

**Proof.** By Theorem 24, we know that $|T| = k_{\min}$. We can obtain the total number of repetitions of $T$ in $G$ by adding the repeats in $A_{2k_{\min}-1}$ and $G \setminus A_{2k_{\min}-1}$. Let $T_A, T_G$ be the portion of $T$ covering $A_{2k_{\min}-1}$ and $G \setminus A_{2k_{\min}-1}$, respectively. As every tour in $T$ traverses level $2k_{\min} - 1$, the total number of repeats of $T_G$ in $G \setminus A_{2k_{\min}-1}$ is at least the total number of repeats of **VD**, and the total number of repeats of $T_A$ in $A_{2k_{\min}-1}$ is at least the total number of repeats of $\mathbf{PA}_{k_{\min}}$; see Theorem 9. Then, $\mathbf{PA}_{k_{\min}} + \mathbf{VD}$ produces a set of tours covering $G$ with minimum number of repeats. ◀

### 6.1.1 $|V(G)|$ is even

▶ **Lemma 28.** *Let $A'_{2k-2}$ be any grid resulting from removing exactly one point from the baseline of $A_{2k-2}$. If $|V(G)|$ is even, then any covering in $G$ that covers $A'_{2k-2}$ with $k-1$ walks performs at least one repetition in $A'_{2k-2}$.*

**Proof.** We know that $|V(A_{2k-2})| = |V(G)| - (k-1)(2k-1)$. If $|V(G)|$ is even, then $|V(A_{2k-2})| \equiv k-1 \mod 2$, implying that $|V(A'_{2k-2})| \equiv k \mod 2$. Now let $N$ be the total number of times that vertices of $A'_{2k-2}$ are traversed by the $k-1$ walks. Notice that any path that starts and ends on the baseline of $A'_{2k-2}$ covers an odd number of points; therefore $N \equiv k-1 \mod 2$. Since $N \geq |V(A'_{2k-2})|$ but they have different parities, $N > |V(A'_{2k-2})|$ and the lemma holds. ◀

▶ **Theorem 29.** *If $|V(G)|$ is even, then $PA_{k_{min}}+VD$ obtains a set of tours covering $G$ with minimum number of repetitions.*

**Proof.** Let $k = k_{\min}$. The number of repeats of $PA_{k_{\min}}+VD$ when $|V(G)|$ is even is $2k^2-2k$. Let $T$ be an optimal solution to MRP and let $d$ be the number of tours in $T$ that do not traverse level $2k-1$. If $d = 0$, then $PA_{k_{\min}}+VD$ obtains an optimal solution for MRP; see Theorem 27. Otherwise, $d = 1$; see Lemma 26. To count the total number of repetitions of $T$ in $G$, we add the repetitions of $T$ in $A_{2k-2}$ and $G \setminus A_{2k-2}$. By Theorem 23, we know the total number of repeats of $T$ in $G$ is at least $2k^2 - 2k - 1$; however this only accounts for the repetitions of $T$ in $G \setminus A_{2k-2}$. As $d = 1$, counting the repeats of $T$ in $A_{2k-2}$ is equivalent to counting the repetitions of $k-1$ walks in $A'_{2k-2}$; see the notation and the results from Lemma 28. As $|V(G)|$ is even, the total number of repeats of $T$ matches the total number of repeats of $PA_{k_{\min}}+VD$. ◀

### 6.1.2 $|V(G)|$ is odd

When $|V(G)|$ is odd and $k_{\min} < \lceil \frac{n_c}{2} \rceil$, we know that $PA_{k_{\min}}+VD$ may not be optimal if exactly one tour in an optimal solution does not reach the level $2k_{\min} - 1$; see Lemma 26 and Theorem 27. We will describe a modified version of **PA** for this case. In the modified version of **PA**, or **PAO**dd, we replace OT$\mathcal{U}$-Covering for **One Tour $\mathcal{U}_{odd}$-Covering**; see Algorithm 6.1. For a given $r$, **PAO** aims to compute a covering of $A_{2r-2}$ with $r-1$ walks and one point. This point will be the last point on the baseline of $A_{2r-2}$; see the red cross in Figure 10. As **PAO** is only applied when $|V(G)|$ is odd, the case described in Figure 10 is complete. Notice that in this scenario, it is guaranteed the existence of a set of $r-1$ walks and one point covering $A_{2k_{\min}-2}$; thus, we can apply the same ideas from Theorem 8 (see Section 3.2) to prove that **PAO** always finds a set of $k_{\min}$ walks covering $A_{2k_{\min}-2}$. As **PAO** covers $A_{2k_{\min}-2}$ without repetitions, it is easy to see that the following theorem holds:

▶ **Theorem 30.** *Let $T$ be a set of tours covering $G$ with minimum number of repeats such that there exists a tour in $T$ that does not traverse level $2k_{min} - 1$. If $|V(G)|$ is odd, then $PAO+VD$ produces a covering of $G$ with minimum number of repeats in $O(|V(G)|)$ time.*

## 6.2 The case $k_{min} = \lceil \frac{n_c}{2} \rceil$

We begin with a simple result that can be established using reasoning analogous to the proof of Theorem 29.

▶ **Theorem 31.** *If $k_{min} = \lceil \frac{n_c}{2} \rceil$ and $n_c$ is even, then the set of tours resulting from $PA_{k_{min}}+VD$ is a covering of $G$ with minimum number of repetitions.*

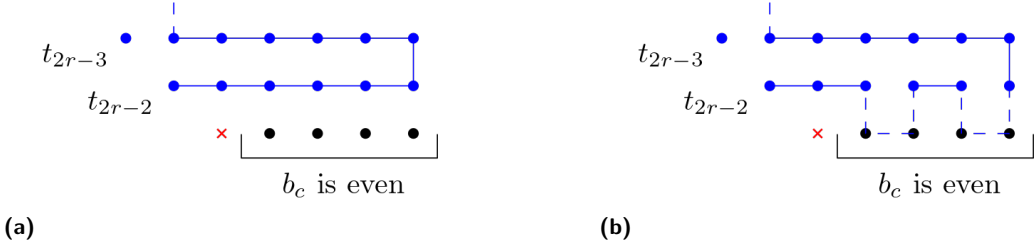

**Figure 10** Final step in OT$\mathcal{U}_{odd}$-Covering when $|V(G)|$ is odd. The red cross indicates the point on the baseline of $A_{2r-2}$ that will not be covered. (a) State of the covering (indicated with the blue color) after applying $\mathcal{U}$-Covering. (b) Modification of the path created with $\mathcal{U}$-Covering. Solid lines are used for edges in the original path, while dashed lines at the bottom are new edges.

**■ Algorithm 6.1 One Tour $\mathcal{U}_{odd}$-Covering** ($OT\mathcal{U}_o$-Covering)

**Require:** $A = AddRow(AddRow(U))$ for some $U \in \mathcal{S}$ with an odd number of rows and columns, and $l \in \mathbb{N}$ being at least the area of $A$. Assume rows in $A$ are numbered from bottom to top. Let $A'$ be the grid resulting from removing the leftmost vertex of the first row in $A$.

**Ensure:** A full covering of $A'$, starting at the leftmost vertex of the third row of $A$, and ending in the leftmost vertex of the second row of $A$.

**1.** Step 1 is similar to step 1 of $\mathcal{U}$-Covering; see Algorithm 3.1.
**2.** Step 2 is similar to step 2 of $\mathcal{U}$-Covering; see Algorithm 3.1.
**3.** If the uncovered portion of $A$ is a rectangle with one row, then finish the covering of $A$ using a comb as detailed in Figure 10.
**4.** Go to step 2.

When $k_{\min} = \lceil \frac{n_c}{2} \rceil$ and $n_c$ is odd, the minimum number of repeats required to cover $G$ can be really high. For instance, the grid in Figure 11 may need many repeats above level $n_c - 1$ when $L$ is the perimeter of the grid. This situation can be easily extended to other grid graphs with an odd number of columns and with a *small* value of $L$. In what follows, we assume $k_{\min} = \lceil \frac{n_c}{2} \rceil$ and $n_c$ is odd, unless otherwise specified.

▶ **Lemma 32.** *Given $A_{n_c-1}$ with an odd number of points on its baseline, consider $\mathcal{P}^*$ as a covering of $A_{n_c-1}$ with minimum number of repeats using $\lceil \frac{n_c}{2} \rceil$ walks starting and ending at on baseline. Let $l_{\min} = \min\{\ell(p_j) : p_j \in \mathcal{P}^*\}$; then the number of repeats of $\mathcal{P}^*$ in $A_{n_c-1}$ is at least $\frac{l_{\min}}{2}$.*

**Proof.** Let $d_k$ be the level $n_c - 2 + k$ of $G$; then $d_1$ represents the baseline of $A_{n_c-1}$. Notice that any $d_k$ contains at most $n_c$ points. However, as $\lceil \frac{n_c}{2} \rceil$ walks are used to cover $A_{n_c-1}$, $n_c$ is odd, and each walk starts and ends on $d_1$, it holds that $n_c + 1$ points are used by $\mathcal{P}^*$ on $d_1$, enforcing one repetition by the Pigeonhole Principle. We generalize this idea by introducing some additional concepts. Notice that any walk can be treated as a directed walk starting and ending on the baseline of $A_{n_c-1}$. We say that a walk goes through the diagonal $d_i$ if it has one edge exiting $d_i$ and one edge entering $d_i$; see Figure 12. Let $e_p(i)$ be the number of edges of walk $p$ exiting and entering $d_i$. If all walks goes through $d_i$, then for any $p \in \mathcal{P}^*$ the number of repetitions in $d_i$ is at least $\frac{e_p(i)}{2}$. Now, let $d_j$ be the last diagonal such that all walks in $\mathcal{P}^*$ go through $d_i$ for all $i \leq j$, and let $p_{\min} \in \mathcal{P}^*$ be a walk such that $\ell(p_{\min}) = l_{\min}$. If $l_{\min} = \sum_{i=1}^{j} e_{p_{\min}}(i)$, then the theorem holds. Otherwise, let $d_k$ with $k \geq j+1$ be the last diagonal such that $p_{\min}$ goes through $d_k$, and let $p$ be a walk that does not go trough $d_j$.

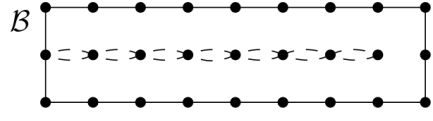

**Figure 11** When $k_{\min} = \lceil \frac{n_c}{2} \rceil$, $n_c$ is odd and $L$ equals the grid perimeter, an optimal solution to MRP can contain many repetitions above the level $n_c - 1$. For visualization purposes, this grid has been rotated.

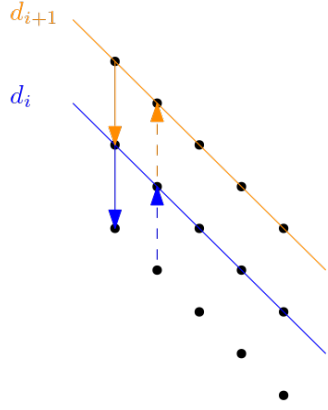

**Figure 12** Example of entering and exiting edges in $d_i$ and $d_{i+1}$. The edges are part of directed walks starting and ending on $d_1$ (the baseline of $A_{n_c-1}$). Solid and dashed lines represent exiting and entering edges, respectively. Each color associate edges with a level.

Then, $\forall z$ with $j < z \leq k$, $p$ performs $\frac{e_{p_{\min}}(z)}{2}$ repetitions at some $d_v$ with $v \leq j$; otherwise $p_{\min}$ would not be of minimum length. Adding all repeats up to $d_k$, we get the theorem. ◄

▶ **Lemma 33.** *Assume $n_c$ is odd and $n_r$ is even. Then, any set of $k = \lceil \frac{n_c}{2} \rceil$ walks covering $A_{n_c-1}$ that start from and return to its baseline performs at least one repeat.*

**Proof.** (By contradiction). Let $P$ be a set of $k$ walks covering $A_{n_c-1}$ with no repeats. As $n_c$ is odd, there exists exactly one walk of length 0 in $P$ covering just one point on the baseline; let $p_s$ be such a walk, where the vertex covered by $p_s$ is at position $s$ from left to right on the baseline of $A_{n_c-1}$. Since walks in $P$ start and end on the baseline of $A_{n_c-1}$, they are clearly disjoint if and only if for any walk starting at point $i$ and ending at point $j > i$ of the baseline, it follows that any walk starting at $i < z < j$ ($z \neq s$) ends at $z < w < j$. Then, $P \setminus \{p_s\}$ can be transformed into $k - 1$ disjoint cycles in $G$ covering exactly $n = n_c n_r - 1$ vertices of $G$; hence $n$ must be even. Then $n_c n_r$ is odd, a contradiction. ◄

For the case considered in this section, we modify OT$\mathcal{U}$-Covering so it can produce a walk that covers a single column. This walk goes up and down covering all the required points of the column. We denote this method as OT$\mathcal{U}_R$-Covering, and despite its simplicity, it is described in Algorithm 6.2 for completeness. We distinguish the version of **PA** using OT$\mathcal{U}_R$-Covering instead of OT$\mathcal{U}$-Covering as **PAR**.

▶ **Lemma 34.** *Let $U = \mathcal{R}(1, b)$ be the last uncovered portion of $A_{n_c-1}$ after applying all steps of **PAR** except for the last one. If $b > 2$ then, all walks of **PAR** covering $A_{n_c-1} \setminus U$ are of maximum length $l$.*

▬ **Algorithm 6.2** One Tour $\mathcal{U}_R$-**Covering** ($OT\mathcal{U}_R$-Covering)

**Require:** $A = \mathcal{R}(1,b)$, and $l \in \mathbb{N}$ being at least two times the perimeter of $A$.
**Ensure:** A covering of A with minimum number of repeats.
  **1.** Return the sequence of vertices of $A$ ordered from bottom to top, and then from top to bottom.

**Proof.** (By contradiction) Let $P = \{p_1, \ldots, p_m\}$ be the set of walks resulting from **PAR** over $A_{n_c-1} \setminus U$; clearly, $P$ is obtained using $\mathcal{U}$-Covering. Let $Z_i$ represent the input grid to the $\mathcal{U}$-Covering algorithm at each step of **PAR**, $p_i$ the path obtained with the algorithm and $R_i = Z_i \setminus p_i$. Assume that there exists $p_j \in P$ such that $\ell(p_j) < l$. Then clearly $R_j$ contains at most one row; otherwise $p_j$ can be increased. Moreover, then $Z_{j+1}$ contains at most three rows, and $R_{j+1}$ at most one, as any $p_i \in P$ covers at least two rows. Then $R_m$ contains at most one row; hence $b \leq 2$, a contradiction. ◀

▶ **Lemma 35.** *Let $P$ be the set of walks resulting from applying **PAR** over $A_{n_c-1}$ with length at most $l$, and $p_m$ the walk with smallest length in $P$. Let $p_{\min}$ be the walk with smallest length in a covering of $A_{n_c-1}$ with minimum number of repeats. Then $\ell(p_m) \leq \ell(p_{\min})$, or $P$ covers $A_{n_c-1}$ optimally with one repeat.*

**Proof.** Recall **PAR** applies $\mathcal{U}$-Covering iteritevely until the last step, when it applies $OT\mathcal{U}_R$-Covering. No repeats are introduced during the application of $\mathcal{U}$-Covering; hence we only need to check the last step of the algorithm. As $n_c$ is odd, $OT\mathcal{U}_R$-Covering is always applied over a rectangle $U = \mathcal{R}(1,b)$. We divide the proof in cases depending on the value of $b$.

Assume $P$ covers $A_{n_c-1}$ with exactly one repeat, then $b = 2$. Note that all walks in $P$ covering $A_{n_c-1} \setminus U$ fully covers an even number of rows (iterative application of Lemma 11). As $b = 2$, then $n_r$ is even, and at least one repeats is required to cover $A_{n_c-1}$; see Lemma 33. Then, $P$ covers $A_{n_c-1}$ with minimum number of repeats.

If $P$ covers $A_{n_c-1}$ with 0 or more than one repeat, then $b = 1$ or $b > 2$, respectively. In the former, the length of the last walk is 0; hence is minimal. In the later, all the previous walks are pairwise disjoint and of maximum length; see Lemma 34. Hence, the last walk obtained with $OT\mathcal{U}_R$-Covering is of minimum possible length among any set of $\frac{n_c+1}{2}$ walks covering $A_{n_c-1}$ with length at most $l$. ◀

Joining the results from Lemma 32 and Lemma 35, we solve the problem:

▶ **Theorem 36.** *If $k_{min} = \lceil \frac{n_c}{2} \rceil$ and $n_c$ is odd, then the set of tours resulting from **PAR+VD** is a covering of $G$ with minimum number of repetitions.*

## 7 Conclusion and future work

In this paper, we consider the problem of computing an optimal covering of a rectangular grid graph with tours of limited length that leave from and return to a base station located at a corner point of the graph. We address two optimization problems: compute a set of tours so that (1) the number of tours is minimized and, (2) the total length, that is, the sum of the lengths of all tours, is minimized. These problems are open for general solid grid graphs. We provide efficient algorithms to solve the rectangular case when each tour is constrained to start at a corner of the grid. Our procedure is fundamentally based on the relationship identified between the two metrics under consideration. First, we present a closed formula for computing the minimum number of tours covering the grid. Interestingly, the algorithm designed for the first criterion achieves an optimal set of tours for the second in all but a few special cases. We identify these cases and modify the algorithm to address them. Overall, we present an approach that efficiently solves both optimization problems in linear time with respect to the grid size. Moreover, we provide a lower bound on the minimum number of repeats produced by any set of tours covering the grid.

We believe that our strategy can be extended to cases where the base station is located at any vertex on the boundary of a rectangular grid graph. Intuitively, the concept of levels could be generalized from lines to triangular regions. Moreover, these problems remain open for solid staircase grid graphs and rectangular grid graphs with holes.

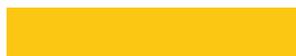